\documentclass[twocolumn]{aastex631}

\usepackage{CJK}
\usepackage{enumitem}
\usepackage{graphbox}
\usepackage{amsmath}
\usepackage{soul}

\begin{document}
\begin{CJK*}{UTF8}{gbsn}
\title{Unveiling the Binary Nature of NGC 2323}

\correspondingauthor{Songmei Qin, Jing Zhong}
\email{qinsongmei@shao.ac.cn, jzhong@shao.ac.cn}

\author[0000-0003-3713-2640]{Songmei, Qin (秦松梅)}
\affiliation{Astrophysics Division, Shanghai Astronomical Observatory, Chinese Academy of Sciences, 80 Nandan Road, Shanghai 200030, China}
\affiliation{School of Astronomy and Space Science, University of Chinese Academy of Sciences, No. 19A, Yuquan Road, Beijing 100049, China}
\affiliation{Institut de Ci\`encies del Cosmos, Universitat de Barcelona (ICCUB), Mart\'i i Franqu\`es 1, 08028 Barcelona, Spain}

\author[0000-0001-5245-0335]{Jing, Zhong (钟靖)}
\affiliation{Astrophysics Division, Shanghai Astronomical Observatory, Chinese Academy of Sciences, 80 Nandan Road, Shanghai 200030, China}

\author [0000-0003-1864-8721]{Tong, Tang (唐通)}
\affiliation{Astrophysics Division, Shanghai Astronomical Observatory, Chinese Academy of Sciences, 80 Nandan Road, Shanghai 200030, China}
\affiliation{School of Astronomy and Space Science, University of Chinese Academy of Sciences, No. 19A, Yuquan Road, Beijing 100049, China}

\author[0000-0002-4986-2408]{Yueyue, Jiang (蒋悦悦)}
\affiliation{Astrophysics Division, Shanghai Astronomical Observatory, Chinese Academy of Sciences, 80 Nandan Road, Shanghai 200030, China}
\affiliation{School of Astronomy and Space Science, University of Chinese Academy of Sciences, No. 19A, Yuquan Road, Beijing 100049, China}

\author[0000-0001-8713-0366]{Long, Wang (王龙)}
\affiliation{School of Physics and Astronomy, Sun Yat-sen University, Daxue Road, Zhuhai, 519082, China}
\affiliation{CSST Science Center for the Guangdong-Hong Kong-Macau Greater Bay Area, Zhuhai, 519082, China}

\author[0000-0003-0349-0079]{Kai, Wu (吴开)}
\affiliation{Department of Physics, School of Mathematics and Physics, Xi'an Jiaotong-Liverpool University, 111 Ren'ai Road, Industrial Park District, Suzhou, Jiangsu, 215123, China}
\affiliation{Department of Mathematical Sciences, University of Liverpool, Liverpool L69 3BX, UK}

\author[0000-0003-4524-9363]{Friedrich Anders}
\affiliation{Departament de F\'isica Qu\`antica i Astrof\'isica (FQA), Universitat de Barcelona (UB), C Mart\'i i Franqu\`es, 1, 08028 Barcelona, Spain}
\affiliation{Institut de Ci\`encies del Cosmos, Universitat de Barcelona (ICCUB), Mart\'i i Franqu\`es 1, 08028 Barcelona, Spain}
\affiliation{Institut d'Estudis Espacials de Catalunya (IEEC), Edifici RDIT, Campus UPC, 08860 Castelldefels (Barcelona), Spain}

\author[0000-0001-9789-7069]{Lola Balaguer-N\'u\~nez}
\affiliation{Departament de F\'isica Qu\`antica i Astrof\'isica (FQA), Universitat de Barcelona (UB), C Mart\'i i Franqu\`es, 1, 08028 Barcelona, Spain}
\affiliation{Institut de Ci\`encies del Cosmos, Universitat de Barcelona (ICCUB), Mart\'i i Franqu\`es 1, 08028 Barcelona, Spain}
\affiliation{Institut d'Estudis Espacials de Catalunya (IEEC), Edifici RDIT, Campus UPC, 08860 Castelldefels (Barcelona), Spain}

\author[0009-0002-4686-8115]{Guimei, Liu (刘桂梅)}
\affiliation{Xinjiang Astronomical Observatory, Chinese Academy of Sciences, Urumqi 830011, China}
\affiliation{School of Astronomy and Space Science, University of Chinese Academy of Sciences, No. 19A, Yuquan Road, Beijing 100049, China}

\author[0000-0003-2021-4818]{Chunyan, Li (李春燕)}
\affiliation{School of Physics and Astronomy, China West Normal University, No. 1 Shida Road, Nanchong 637002, China}

\author{Jinliang, Hou (侯金良)}
\affiliation{Astrophysics Division, Shanghai Astronomical Observatory, Chinese Academy of Sciences, 80 Nandan Road, Shanghai 200030, China}
\affiliation{School of Astronomy and Space Science, University of Chinese Academy of Sciences, No. 19A, Yuquan Road, Beijing 100049, China}

\author[0000-0002-4907-9720]{Li, Chen (陈力)}
\affiliation{Astrophysics Division, Shanghai Astronomical Observatory, Chinese Academy of Sciences, 80 Nandan Road, Shanghai 200030, China}
\affiliation{School of Astronomy and Space Science, University of Chinese Academy of Sciences, No. 19A, Yuquan Road, Beijing 100049, China}



\begin{abstract}

As a well-known open cluster, NGC 2323 (also called M50) has been widely investigated for over a hundred years and has always been considered a classical single cluster. In this work, with the help of {\it Gaia} DR3, we study the binary structure nature of this cluster. Although indistinguishable in the spatial space, the small but undeniable difference in the proper motion indicates that they may be two individual clusters. After investigating the properties of the two clusters, it is found that they have very close positions (three-dimensional $\Delta$pos = 12.3~pc，$\sigma_{\Delta \mathrm{pos}} = 3.4$~pc) and similar tangential velocities (two-dimensional $\Delta$V = 2.2~km~s$^{-1}$， $\sigma_{\Delta \mathrm{V}} = 0.02$~km~s$^{-1}$), indicating the existence of their physical association. Moreover, the best isochrone fitting ages of the two clusters are the same (158~Myr), further proving their possibly common origin. To comprehensively understand the formation and evolution of this binary cluster, we employ the PETAR $N$-body code to trace back their birthplace and deduce their dynamical evolutionary fate. With observational mean cluster properties, the simulations suggest that they may form together, and then orbit each other as a binary cluster for over 200~Myr. After that, because of their gradual mass loss, the two clusters will eventually separate and evolve into two independent clusters. Meanwhile, the numerical $N$-body simulation suggests that the less massive cluster is unlikely to be the cluster tidal tails created by the differential rotation of the Milky Way.

\end{abstract}



\section{Introduction}
\label{sec: intro}

Open Clusters (OCs) are born in giant molecular clouds \citep{2003ARA&A..41...57L}, most of which are isolated stellar systems \citep{2017A&A...600A.106C}, and in some cases also formed in groups, such as binary clusters or higher-order systems \citep{2016MNRAS.455.3126C}. Observed diverse OC groups or complexes provide an ideal laboratory for studying the physical principles underlying the hierarchical structure of star-forming regions \citep{1981MNRAS.194..809L,2003MNRAS.343..413B,2022ApJ...931..156P}. At the same time, exploring OC groups can help us to understand the extension of the hierarchical star formation process to larger galactic scales \citep{1996ApJ...466..802E}.

\citet{2009A&A...500L..13D} recapitulated several formation channels of binary clusters. Co-formation scenario suggested that binary clusters are genetic or primordial pairs, which are formed from the same molecular cloud and share the same velocity, age, and metallicity \citep{1997AJ....113..249F,2004ApJ...602..730B}, such as the extensively studied $h$ and $\chi$ Persei pair (NGC 869/NGC 884) \citep{2002ApJ...576..880S,2019A&A...624A..34Z}. Another scenario is that the stellar winds or supernova shocks would induce the collapse of the cloud and the sequential formation of a companion cluster \citep{1995ApJ...440..666B}, resulting in different ages but similar velocities of binary clusters. The tidal capture mechanism proposed that two clusters with small relative velocities may be tidally captured and even merge when encountering each other \citep{2009A&A...500L..13D}. They are possibly not primordial; therefore, significant age and chemical composition differences can be observed, such as the recently discovered colliding binary clusters IC 4665 and Collinder 35 \citep{2022MNRAS.511L...1P}. 

Through $N$-body simulation, \citet{2010ApJ...719..104D} investigated the evolution of primordial binary clusters, including mergers, shredded secondaries, and separated twins, while the evolution fate is associated with the initial orbital elements and mass ratio of the cluster pairs. They also demonstrated that primordial binary clusters are not stable for a long time, which is consistent with the observational results of \citet{2021ARep...65..755C,2022Univ....8..113C}. The fact that primordial binary clusters are prone to disintegration or merger due to dynamic evolution, which leads to an apparent lack of binary clusters in the Milky Way \citep{1995A&A...302...86S,2009A&A...500L..13D}.

Before the {\it Gaia} era, reliable cluster member selection was fairly difficult due to the limited astrometric precision, which made identifying and exploring binary star clusters difficult. Based on the distance separation criterion ($\le 30$~pc), \citet{2009A&A...500L..13D} selected 34 pairs with WEBDA\footnote{The Open Cluster Database: \url{https://webda.physics.muni.cz/}} and 27 pairs with NCOVOCC\footnote{The New Optically Visible Open Clusters and Candidates catalog: \url{http://www.astro.iag.usp.br/~wilton/}} and obtained the binary cluster fraction of 12\%, which is comparable to that in the Magellanic Clouds. By adding kinematic information, \citet{2017A&A...600A.106C} applied the Friends-of-Friends algorithm with linking length 100~pc in spatial space and 10~km~s$^{-1}$ in velocity space, and detected 19 groups including 14 pairs.

{\it Gaia} \citep{2018A&A...616A...1G,2021A&A...649A...1G,2023A&A...674A...1G} presents unprecedented high-precision five astrometric parameters ($\alpha$, $\delta$, $\varpi$, $\mu_{\alpha}^*$, $\mu_{\delta}$) and three-band photometry ($G$, $G_{\rm BP}$ and $G_{\rm RP}$). Radial velocities data are also in the released catalog \citep{2019A&A...622A.205K,2023A&A...674A...5K}. As one of the most successful and ambitious projects, {\it Gaia} enabled the increasing discovery of new OCs \citep{2018A&A...618A..93C,2020A&A...640A...1C,2019ApJS..245...32L,2020A&A...635A..45C,2022A&A...661A.118C,2021RAA....21...45Q,2023ApJS..265...12Q,2022ApJS..260....8H,2021A&A...646A.104H,2023A&A...673A.114H}, promoting the burgeoning exploration of binary clusters \citep{2019A&A...623C...2S,2021ARep...65..755C,2022MNRAS.510.5695A,2022MNRAS.511L...1P,2022A&A...666A..75S,2023ApJS..265...12Q,2023MNRAS.521.1399C}. Particularly, \citet{2023ApJS..265...12Q} discovered 324 clusters with data slicing method based on {\it Gaia} Data Release 3 (DR3), of which 101 are new clusters. As the most complete open cluster sample in the solar neighborhood to date, a total of 19 binary clusters and three triple clusters have been identified, with spatial separation of less than 20~pc, velocity difference of less than 5~km~s$^{-1}$, and similar ages.

Based on {\it Gaia} DR3, we serendipitously found that the famous cluster NGC 2323 (M50) is, in its nature, probably a binary cluster. NGC 2323, located in the Monoceros region of the Milky Way, was first discovered by Giovanni Cassini before 1709 and compiled in the Messier catalog by Charles Messier later. NGC 2323 was widely perceived as an isolated star cluster in literature. Early studies of NGC 2323 presented its distance varying from 500 to 1230~pc \citep{1930LicOB..14..154T,1930fcsm.book.....S,1931AnLun...2....1C,1941ApJ....94...55C,1961PUSNO..17..343H,2003AJ....126.1402K} and age ranging from 60 to 180~Myr \citep{1969Ap&SS...3..123B,1998A&AS..128..131C,2003AJ....126.1402K,2012AstL...38...74F,2013A&A...558A..53K,2017SerAJ.194...59A}. Since the {\it Gaia} data released, NGC 2323 has still been identified as a single OC in the most updated OC catalogues \citep{2018A&A...618A..93C,2020A&A...640A...1C,sim_207_2019,2023A&A...673A.114H}. Incorporating {\it Gaia} Data Release 2 and UBV photometric data from the ground-based instrument, \citet{2019JApA...40...33O} estimated the metallicity (Z = 0.012) of NGC 2323 from the $\delta (U-B)$ technique.

\begin{figure*}
   \centering
   \includegraphics[width=\textwidth, angle=0]{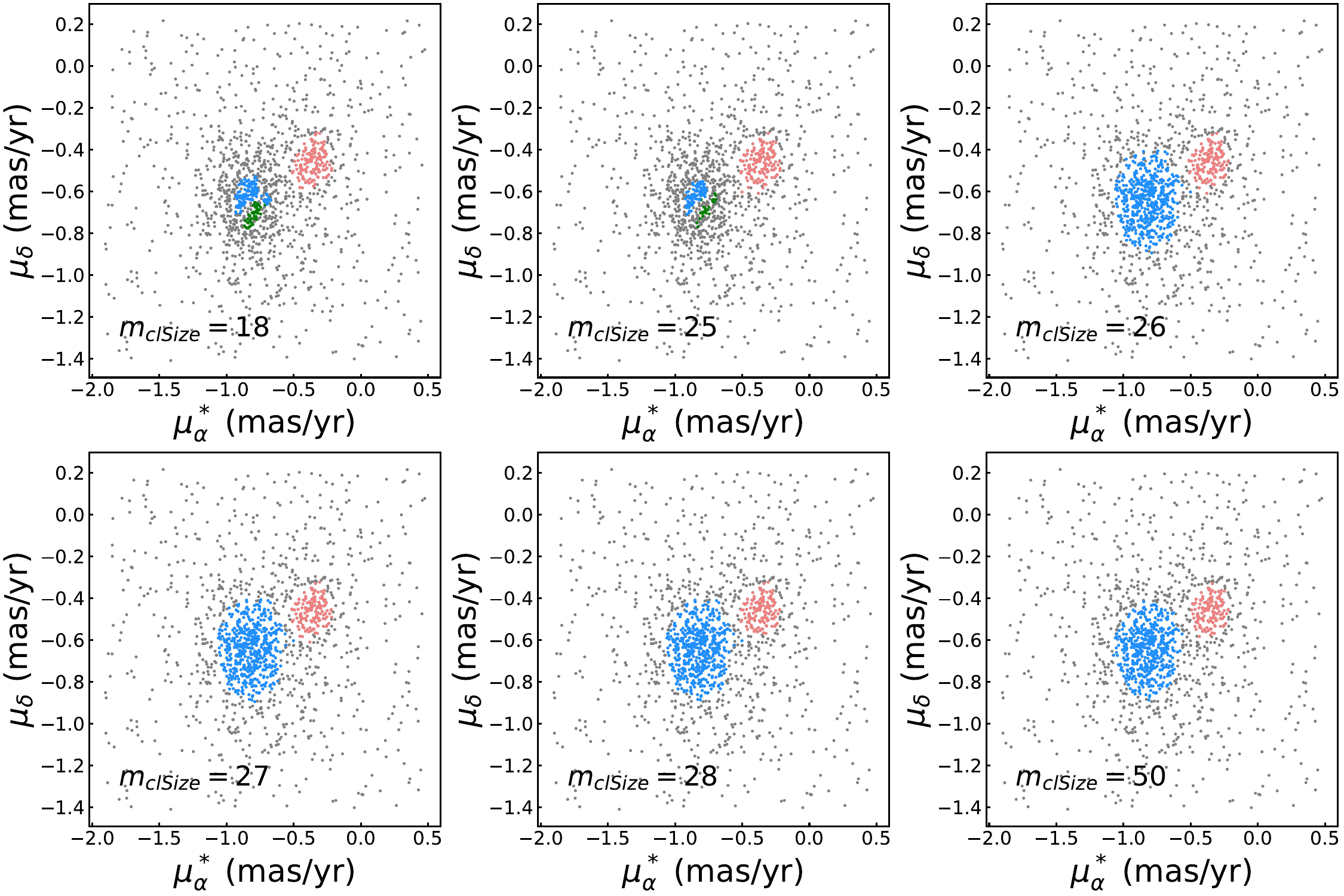}
   \caption{The proper motion ($\mu_{\alpha}^*$, $\mu_{\delta}$) distributions of the clustering results by using HDBSCAN with different min\_cluster\_size ($m_{clSize}$) parameters. The grey dots represent the field stars, and other color-coded dots are the cluster members identified by HDBSCAN.}
   \label{fig: clustering}
\end{figure*}

However, with the help of precise astrometric and photometric {\it Gaia} data, we can now unveil the nature of NGC 2323: it is probably a pair of primordial binary clusters. In Section~\ref{sec: ana}, we confirm the true existence of the binary structures in NGC~2323 and investigate their properties. In Section~\ref{sec: o-e}, we discuss the possible origin of the binary structures, including binary cluster and tidal tail scenarios. A summary of this work is given in Section~\ref{sec: sum}.

\section{Discovery of the binary structure} \label{sec: ana}

By employing the Hierarchical Density-Based Spatial Clustering of Applications with Noise (HDBSCAN) algorithm, \citet{2023A&A...673A.114H} identified 7200 clusters with {\it Gaia} DR3, while the latest property parameters of NGC 2323 ($\alpha$, $\delta$, $\mu_{\alpha}^{*}$, $s_{\mu_{\alpha}^{*}}$, $\mu_{\delta}$, $s_{\mu_{\delta}}$, $\varpi$, $s_{\varpi}$, $R_{\rm tot}$) were included in their catalog. To study the binary structure in the NGC~2323 region, we retrieved the member stars referring to those parameters of NGC 2323 in \citet{2023A&A...673A.114H} and then re-investigate the properties of this star cluster. The selection criteria of astrometric and photometric data from {\it Gaia} Archive\footnote{\url{https://gea.esac.esa.int/archive/}} are as follows:
\begin{enumerate}[itemsep=0.1pt, parsep=0.2pt, topsep=0.2pt]
\item Centering on the approximate central position ($\alpha$, $\delta$) with a radius of $R_{\rm tot}$.
\item Proper motion selection: $\mu_{\alpha}^{*} \pm 5 \cdot s_{\mu_{\alpha}^{*}}$; $\mu_{\delta} \pm 5\cdot s_{\mu_{\delta}}$.
\item Parallax selection: $\varpi \pm 3 \cdot s_{\varpi}$.
\item $G < 20$~mag.
\end{enumerate}

\begin{table*}[htbp]
\begin{center}
\scriptsize
\setlength{\tabcolsep}{3pt}
\setlength{\belowcaptionskip}{0.01cm} 
\renewcommand{\arraystretch}{1.5}
\caption{\centering The property parameters of NGC 2323-a, NGC 2323-b, and NGC 2323-HR.}\label{tab: para}
\begin{tabular}{cccccccccccccccc}
\hline
Cluster$^{\dagger}$   & Number & $\alpha$ & $s_{\alpha}$ & $\delta$ & $s_{\delta}$ &$\varpi$& $s_{\varpi}$ & $d$ & $s_{d}$& $\mu_{\alpha}^*$& $s_{\mu_{\alpha}^{*}}$ & $\mu_{\delta}$  &$s_{\mu_{\delta}}$ & $RV$        & $e_{RV_{\rm_{OC}}}$ \\
           &                   & deg      & deg          & deg      & deg          & mas    & mas          & kpc & kpc    & mas yr$^{-1}$   & mas yr$^{-1}$          & mas yr$^{-1}$   & mas yr$^{-1}$     & km s$^{-1}$ & km s$^{-1}$  \\
\hline
NGC 2323-a & 519              & 105.733  & 0.21  & -8.359  & 0.21 & 1.043 & 0.002 & 0.954$^{\ast}$ & 0.001 & -0.823  & 0.11    & -0.646    & 0.11    & 6.33   & 0.65 \\
NGC 2323-b & 139              & 105.577  & 0.23  & -8.374  & 0.24 & 1.022 & 0.005 & 0.966$^{\ast}$ & 0.003 & -0.363  & 0.07    & -0.461    & 0.06    & 6.55   & 1.67 \\
NGC 2323-HR& 788              & 105.705  & --    & -8.355  & --   & 1.006 & 0.05 & 0.956$^{\star}$ & --    & -0.717  & 0.24    & -0.602    & 0.16    & 5.08   & 1.28 \\
\hline
\end{tabular}
\end{center}
\footnotesize{$\dagger$ NGC 2323-a and NGC 2323-b refer to the binary cluster we identified in this work, and NGC 2323-HR represents the cluster cataloged in \citet{2023A&A...673A.114H}.}\\
\footnotesize{$\ast$ The mean Bayesian distance obtained in this work.}\\
\footnotesize{$\star$ The photometric distance provided in \citet{2023A&A...673A.114H}.}\\
\end{table*}

\begin{figure}
   \centering
   \includegraphics[width=0.45\textwidth, angle=0]{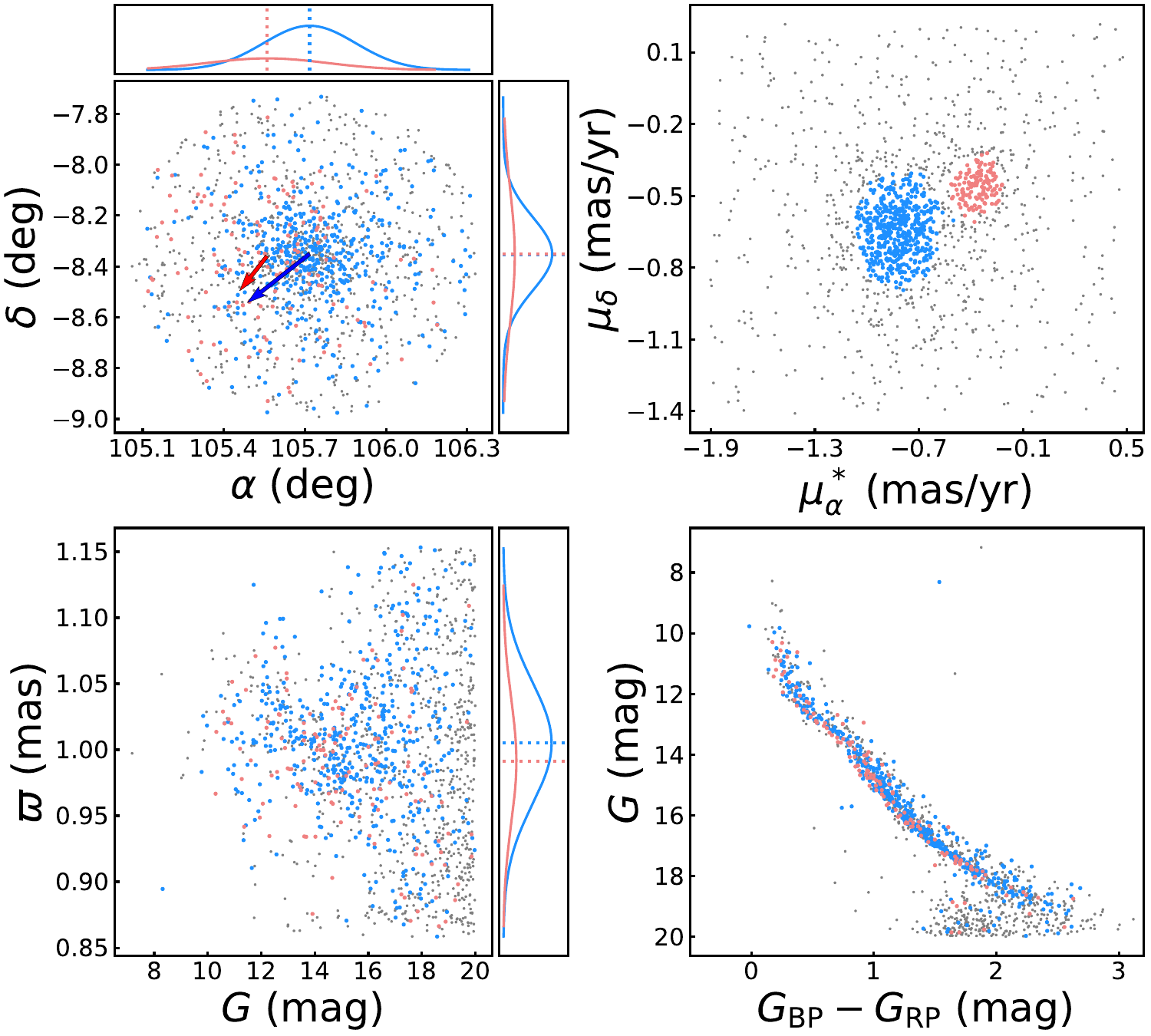}
   \caption{Spatial ($\alpha$, $\delta$), proper motion ($\mu_{\alpha}^*$, $\mu_{\delta}$), magnitude-parallax ($G$, $\varpi$), color-magnitude ($G_{\rm BP}-G_{\rm RP}$, $G$) diagrams of the clustering result by using HDBSCAN. The dodger blue and light coral points refer to the separated cluster members of NGC 2323-a and NGC 2323-b, and the grey points represent the discarded field stars. The blue and red arrows are the average proper motions of the two clusters. The corresponding fitted Gaussian profiles of $\alpha$, $\delta$, and $\varpi$ are added on the left panels for the two clusters, with dotted lines indicating the fitted mean values.}  
   \label{fig: member}
\end{figure}

After excluding foreground and background stars using the above filters, the sample exhibits sole concentration in 2D spatial space and one peak in parallax distribution as shown in the left panels of Figure~\ref{fig: member}. Outwardly, it seems like the symbol of one cluster. However, we can perceive two adjacent but distinct concentrations in the proper motion distribution, which illustrates the existence of two `clusters' with different kinematics in the location of NGC 2323. With the initial sample, we tried min\_cluster\_size ($m_{clSize}$) from 2 to 60 (step = 1) of HDBSCAN algorithm \citep{mcinnes2017hdbscan} in the proper motion space, and the true cluster number and the cluster member results remain stable when $m_{clSize}$ reaches 27, as shown in Figure~\ref{fig: clustering}. Hence, we adopted $m_{clSize}$ = 27 as an optimal parameter to perform the HDBSCAN \citep{mcinnes2017hdbscan} clustering process. As a result, we identified two `clusters' named NGC 2323-a and NGC 2323-b and obtained 519 and 139 members, respectively. In the upper left panel of Figure~\ref{fig: member},  it can be seen that these two `clusters' are located very close in the projection spatial space (NGC 2323-a has more members and is more concentrated than NGC 2323-b) and roughly have a similar tangential velocity direction. 

Based on their members, we provided the cluster properties for NGC 2323-a and NGC 2323-b as listed in Table~\ref{tab: para}. We derived the mean and standard deviation values of the positions and proper motions for the two `clusters'. Then we calculated the average radial velocity and corresponding uncertainty for each cluster using the same procedure described by \citet{2018A&A...619A.155S,2022A&A...658A..14C}. The average radial velocity is obtained using the following formula:
\begin{equation}\label{eq:meanrv}
RV_{\rm{OC}} = \frac{\sum_i RV_{i} \times w_i}{\sum_i w_i},
\end{equation}
where $w_i$ is defined as $w_i = 1/ (\epsilon_{i})^{2}$. $RV_{i}$ and $\epsilon_{i}$ are the radial velocity and corresponding error provided by {\it Gaia} DR3 for each member. The $RV_{\rm{OC}}$ uncertainty is defined as the maximum of $\sigma_{RV_{\rm{OC}}} / \sqrt{N}$ and $I / \sqrt{N}$, and the $RV_{\rm{OC}}$ weighted standard deviation $\sigma_{RV_{\rm{OC}}}$ is defined as:

\begin{footnotesize}
\begin{equation}\label{eq:estd}
\sigma_{RV_{\rm{OC}}}=\sqrt{\frac{\sum_i w_i}{(\sum_i w_i)^2-\sum_i w_i^2}\times\sum_i w_i\times(RV_{i}-RV_{OC})^2},
\end{equation}
\end{footnotesize}
and the internal error of $RV_{\rm{OC}}$ is
\begin{equation}\label{eq:eint}
I = \frac{\sum_i w_{i} \times \epsilon_{i}}{\sum_i w_i}.
\end{equation}
To discard stars with discrepant radial velocities with respect to $RV_{\rm{OC}}$ because of binaries, we iteratively excluded the stars with velocities outside the $RV_{\rm{OC}} \pm 3 \times \sigma_{RV_{\rm{OC}}}$. We corrected parallax zeropoint offsets for all the members \citep{2021A&A...649A...4L} and obtained the mean parallaxes and standard deviation values. Then we assumed that the cluster is very small compared to its distance and adapted an exponential prior and a Gaussian distribution of parallax \citep{2015PASP..127..994B} to determine the Bayesian distances of $0.954 \pm 0.001$~kpc for NGC 2323-a and $0.966 \pm 0.003$~kpc for NGC 2323-b.

\begin{figure*}
   \centering
   \includegraphics[width=0.8\textwidth, angle=0]{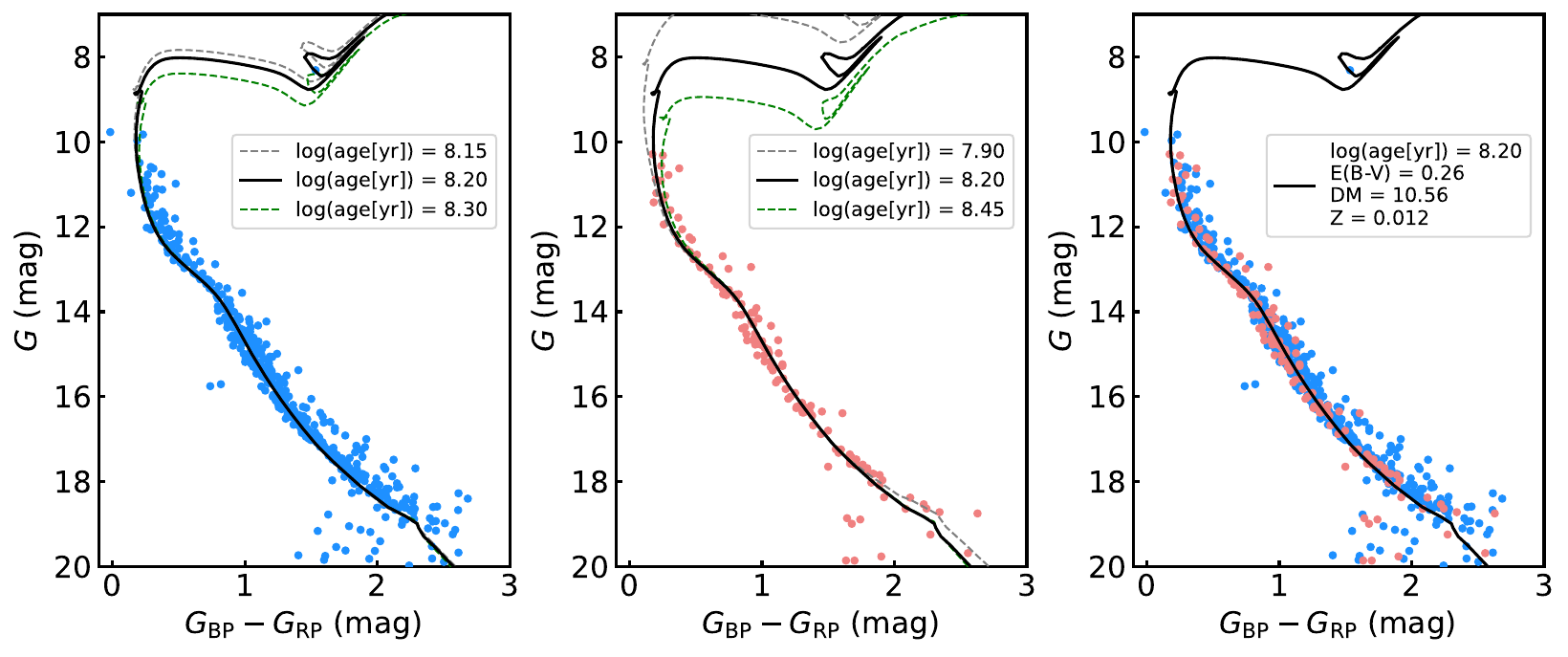}
   \caption{Isochrone fitting of two clusters. The dodger blue and light coral dots refer to NGC 2323-a and NGC 2323-b members identified in this work respectively. The black solid lines indicate the best-fitting isochrone, while the grey and green dashed lines denote the isochrones with the lower and upper limits of visually fitted age.}
   \label{fig: age}  
\end{figure*}

\begin{table*}[htbp]
\setlength{\tabcolsep}{3.4pt}
\setlength{\belowcaptionskip}{0.3cm}
\renewcommand{\arraystretch}{1.5}
\small
\caption{The property parameters of NGC 2323-a and NGC 2323-b.}\label{tab: Gal-para}
\begin{tabular}{cccccccccc}
\hline
Cluster    & X       & Y      & Z    & U           & V           & W           & log(age[yr]) & DM    & E(B-V)    \\
           & pc      & pc     & pc   & km s$^{-1}$ & km s$^{-1}$ & km s$^{-1}$ &    & mag   & mag\\
\hline
NGC 2323-a & -8712.4 & -634.2 & -4.8 & 5.9         & 230.2       & 2.2         & 8.20     & 10.56 & 0.26  \\
NGC 2323-b & -8722.0 & -641.4 & -7.5 & 5.9         & 229.9       & 4.4         & 8.20     & 10.56 & 0.26  \\
\hline
\end{tabular}
\end{table*}

To determine other parameters of the two `clusters' through the color-magnitude diagram (CMD), we downloaded a set of PARSEC isochrones \citep{2017ApJ...835...77M} with {\it Gaia} photometric system \citep{2021yCat..36490003R} from CMD 3.7\footnote{\url{http://stev.oapd.inaf.it/cgi-bin/cmd}} to perform the isochrone fitting. The isochrones used for fitting have a grid of logarithm ages from 7.0 to 9.0 with an interval of 0.05 and an abundance of  $Z$ = 0.012 \citep{2019JApA...40...33O}. Then, we carefully inspected the matching of the isochrones to the characteristic regions, such as the upper main sequence, the turn-off point, and the red giant in the CMDs. We adopted the formula $A_{\rm{G}} = 2.74 \times E(B-V)$ and $E(BP-RP)=1.339 \times E(B-V)$ \citep{2018MNRAS.479L.102C} to calculate the reddening values. Finally, by traversing the grid of isochrones in the CMDs, we used the `blue edge' for visual isochrone fitting and obtained the optimal fitted parameters for NGC 2323-a and NGC 2323-b. The age parameters from isochrone fitting are $158^{+41}_{-17}$~Myr for NGC~2323-a and $158^{+123}_{-79}$~Myr for NGC~2323-b. Both clusters have the same optimal fitted result with log(age[yr]) = 8.20, distance modulus DM = 10.56~mag, and reddening $E(B-V)$ = 0.26~mag as shown in Figure~\ref{fig: age}. It is noticed that the mean distances (both are 938~pc) from isochrone fitting are slightly smaller than Bayesian distances. 

\section{Binary Structure Origin}\label{sec: o-e}

Since star clusters form in hierarchically collapsing gas clouds \citep{2019A&A...626A..17C,2020SSRv..216...64K,2022ApJ...931..156P,2024arXiv240416923Z}, the two different clumps in the Figure~\ref{fig: member} (close positions and similar yet distinctive velocities) illustrate the existence of two clusters in the NGC 2323 region, or the less massive `cluster' is possibly to be the tidal tail or sub-structure of the open cluster NGC~2323. Hence, we used $N$-body simulation to discuss the possible origin of this binary structure.

\begin{table*}
\centering
\small
\setlength{\tabcolsep}{4pt}
\setlength{\belowcaptionskip}{0.001cm}
\renewcommand{\arraystretch}{1.5}
\caption{Initial parameters for McLuster.}\label{tab: sim}
\begin{tabular}{cccccccccccccc}
\hline
Cluster    & $N_{\rm t}$ & PDMF               & $M_{\rm min}^{*}$ & $\alpha_{\rm l}$ & $M_{\rm tra}^{*}$ & $\alpha_{\rm h}$ & $M_{\rm max}^{*}$ & Density profile & $W_{0}$ & $R_{\rm hm}$& $S$  & $Q$ & $f_{\rm b}$ \\
           &             &                    & $\rm M_\odot$         &                  & $\rm M_\odot$         &                  & $\rm M_\odot$         &                 &         & pc          &      &     &             \\
\hline
NGC 2323-a & 1208        & two-part power-law & 0.08              & -0.54            & 0.98              & -3.18            & 4.26              & King            & 3       & 2.99        & 0, 1 & 0.5 & 30\%  \\
NGC 2323-b & 415         & two-part power-law & 0.08              & -0.98            & 1.26              & -2.80            & 3.44              & King            & 1       & 4.14        & 0, 1 & 0.5 & 30\%  \\
\hline
\end{tabular}
\end{table*}

\subsection{Binary cluster}\label{sub: bc}

We calculated the mean positions and velocities in Cartesian Galactic coordinates with \textbf{Astropy} package\footnote{\url{https://docs.astropy.org/en/stable/}} \citep{astropy:2013,astropy:2018,astropy:2022} for the two clusters, as shown in Table~\ref{tab: Gal-para}. We adopted [10, 235, 7]~km s$^{-1}$ and 8.0~kpc as the galactocentric velocities of the Sun and the distance from the Sun to the Galactic center, respectively \citep{Bovy2015}. The mean central positions of the two clusters are close with three-dimensional $\Delta \mathrm{pos} = 12.3$~pc $<20$~pc ($\sigma_{\Delta \mathrm{pos}} = 3.4$~pc), and the tangential velocity difference between the two clusters is $\Delta \mathrm{V} = 2.2$~km~s$^{-1}$ $< 5$~km~s$^{-1}$ ($\sigma_{\Delta \mathrm{V}} = 0.02$~km~s$^{-1}$), which suggest that they may be physically associated binary cluster \citep{1995A&A...302...86S,2019A&A...623C...2S}. In addition, the distribution of the two clusters in the color-magnitude diagram (Figure~\ref{fig: age}) indicates that they are coeval.

To trace back the origin of the binary cluster and explore their future dynamical evolution, we exploited numerical $N$-body simulations for in-depth analysis. At first, we derived the basic parameters from the observational data as input parameters for McLuster\footnote{\url{https://github.com/ahwkuepper/mcluster}} \citep{2011MNRAS.417.2300K}, which can be used to generate the two clusters for the following $N$-body computations. Then, we employed the new high-performance $N$-body code called PETAR\footnote{\url{https://github.com/lwang-astro/PeTar}} \citep{2020MNRAS.497..536W} to conduct numerical simulations of the binary cluster. Finally, based on the result of PETAR, we investigated the kinematic states of the binary cluster at different times.

The star cluster initial model generator McLuster \citep{2011MNRAS.417.2300K} can generate star clusters with various options and parameters. To create a mock cluster, we need to set the input parameters, which include a dimensionless value that specifies the King model concentration ($W_{\rm 0}$), half-mass radius ($R_{\rm hm}$), degree of mass segregation ($S$), number of cluster members ($N$) or the mass of cluster ($M$),  mass function parameters (user defined two-segment power law function, the slope of low mass part ($\alpha_{\rm l}$), the slope of high mass part ($\alpha_{\rm h}$), minimum star mass ($M^{*}_{\rm min}$), transition mass between two segments ($M^{*}_{\rm tra}$), maximum star mass ($M^{*}_{\rm max}$)), virial ratio ($Q$), and the binary fraction ($f_{\rm b}$). 

Based on the observational data of the binary cluster, we performed the radial density profile fitting using the king model \citep{1962AJ.....67..471K} and obtained their concentration parameters $c = {r_{\rm t}}/{r_{\rm c}}$, where $r_{\rm t}$ refers to the tidal radius and $r_{\rm c}$ refers to the core radius of clusters. According to the correspondence between $c$ and another concentration parameter $W_{\rm 0}$ \citep{1966AJ.....71...64K}, we derived $W_{\rm 0}$ equals to 3 and 1 for the two clusters respectively. We applied the degree of mass segregation $S$ = 0, 1 for no and maximum mass segregation to those density profiles. Then, we utilized the best-fitting isochrone in Section~\ref{sec: ana} to acquire the mass of cluster members \citep{2024arXiv240511853J} with $G < 18$~mag\footnote{It should be noted that leaving out the faint stars would bias the half mass radius of the two clusters.} and estimated the half-mass radius for each cluster. At the same time, we respectively derived the low mass slope $\alpha_{\rm l}$, high mass slope $\alpha_{\rm h}$ and transition mass between two segments $M^{*}_{\rm tra}$ parameters of each cluster using a two-part power-law form \citep{2023MNRAS.525.2315A}. It is noted that the two white dwarfs \citep{2016ApJ...818...84C} in NGC 2323 were not included in our simulation, which would have little effect on our simulations. The total number ($N_{\rm t\_a} = 1208$, $N_{\rm t\_b} = 415$) and mass ($M_{\rm t\_a} = 827$~$\rm M_\odot$, $M_{\rm t\_b} = 250$~$\rm M_\odot$) of the two clusters including unseen stars were estimated by integrating the extrapolated observed mass function down to the lower mass limit (0.08~$\rm M_\odot$). To derive the dynamic relaxation time of each cluster, we used the formula provided by \citet{1987gady.book.....B} for estimation:
\begin{equation}
t_{\rm relax} = \frac{0.14\,N\,\sqrt{r_{\rm h}^{3} / G\,M}}{\mathrm{ln} (0.4\,N)},
\end{equation}
where $G$ is the gravitational constant, $N$ is the total member number, $M$ is the total cluster mass, and $r_{\rm h}$ is the half-mass radius. The dynamic relaxation ages of NGC 2323-a and NGC 2323-b are 69~Myr and 85~Myr, respectively. Considering that the ages of both clusters are older than their relaxation ages, the binary cluster has evolved to a relaxation state. \citet{2003AJ....126.1402K} also suggested that NGC 2323 is somewhat dynamically relaxed ($t_{\rm relaxed} = 102$~Myr, $t_{\rm age} = 130$~Myr). Hence, we assumed they are both in virial equilibrium ($Q = 0.5$). 
The binary fraction of our mock clusters is adopted as a typical fraction value of 30\% \citep{2022ApJ...930...44L}. Based on the above parameters listed in Table~\ref{tab: sim}, we utilized McLuster \citep{2011MNRAS.417.2300K} code to simulate the two clusters. It is noted that the total mass discrepancies between the simulated star clusters ($S$ = 0, 1) and the actual clusters (extrapolated observed mass function) are about 3.1\% and 4.5\%, which indicates that the mock clusters have a high similarity to the observational clusters in total mass.

\begin{figure*}
   \centering
   \includegraphics[width=0.9\textwidth, angle=0]{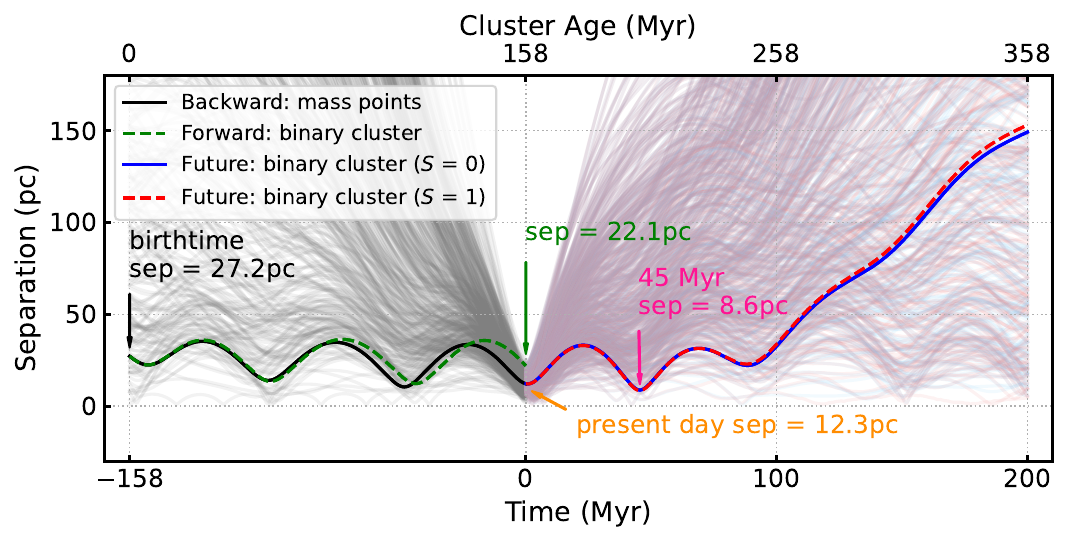}
   \caption{The backward (black line), forward (green dashed line), and future orbital separation (blue line with mass segregation degree $S$ = 0 and red dashed line with mass segregation degree $S$ = 1) simulated by using PETAR with mean property parameters for the binary cluster. The light-shaded curves correspond to the orbits simulated with MCMC sampling over the measured uncertainties in 3D positions and velocities.}
   \label{fig: orb}  
\end{figure*}

It is difficult to simulate the dynamical evolution of the binary cluster from their birth time because we cannot directly observe the initial state of the clusters, such as the initial total mass, the initial mass function, the initial mass segregation, the initial density profile, the initial velocity distribution, and the initial position in the Milky Way. Hence, we treated the two mocked clusters as two mass points to integrate their orbits in the past. The gravitational interaction between them in the Milky Way was considered with PETAR \citep{2020MNRAS.497..536W}. However, the stellar evolution option was turned off during the simulation. We used the positions and velocities of the binary cluster in Cartesian Galactic coordinates at present-day (see Section~\ref{sec: ana}) to trace back the orbits of the binary cluster to their birthplaces. Finally, PETAR provided the positions and velocities of each mass point with a time interval of 0.5~Myr. Figure~\ref{fig: orb} shows their orbital separation (black line) from the birth time (-158~Myr) to the present day (0~Myr). Through the simplified simulation with cluster mean property parameters, it is found that the two clusters were physically separated by a distance of 27.2~pc at their birthplaces, which is smaller than the typical size of the molecular cloud (a hundred~pc, \citet{2015ARA&A..53..583H}). This indicates that they may have been born from the same molecular cloud.

To check if the simplified assumptions in the backward simulation are justified, we applied the forward simulation (including stellar formation, etc.) of the binary cluster from 158 Myr ago to the present day. However, the initial cluster mass, the initial mass function, the initial mass segregation, and the initial density profile are unknown. For simplicity, we assumed that the initial mass and density distributions are the same as the distributions at present. Figure~\ref{fig: orb} illustrates the backward and forward orbital separation distributions. The separation between the two clusters obtained from the forward simulation is 22.1~pc, showing a 9.8~pc discrepancy compared to the observation. The similar pattern of oscillatory motion in the mutual separation of the clusters indicates the self-consistency of the simplified assumptions in the backward simulation. Hence, we regard it rational to trace back the orbit of the binary cluster under the simplified assumption of mass points using PETAR.

\begin{figure*}
   \centering
   \includegraphics[width=\textwidth, angle=0]{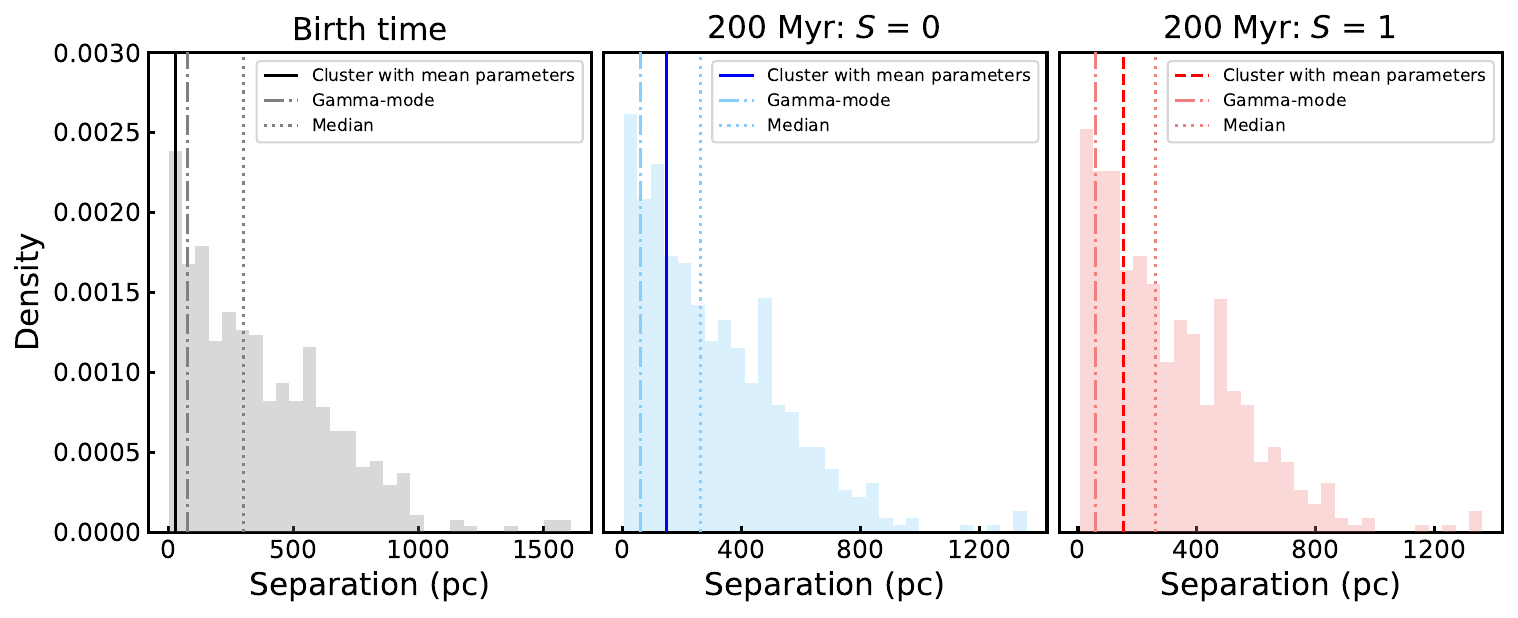}
   \caption{The histogram of orbital separation at birth time and 200 Myr (cluster with mass segregation S = 0 and 1). The black line, blue line, and red dashed line represent the orbital separation for the two clusters with mean property parameters (correspond to lines in Figure~\ref{fig: orb}), and the dash-dot lines refer to the mode values of orbital separations for clusters fitted with gamma function. The dotted lines refer to the median values of orbital separations for clusters with MCMC sampling parameters. The dash-dot lines represent the mode values from the gamma fitting of the density distributions.}
   \label{fig: sep-hist}  
\end{figure*}

Using the mock clusters as input samples, we simulated the future dynamical evolution state of the binary cluster with PETAR \citep{2020MNRAS.497..536W}, which is quite different from the backward simulation under the simplifying assumption of point mass. The simulation started from the gas-free phase, and only particle-like stellar components were contained during the modeling. The evolution of the binary cluster is controlled by the combination of single stellar evolution (SSE) and binary stellar evolution (BSE) \citep{Hurley2000, Hurley2002, Banerjee2020}, stellar interaction, and the external Galactic potential \citep{Bovy2015}. Since the maximum mass of the mocked cluster members is 4.26~$\rm M_\odot$, and the evolution of 99.8\% stars within 158~Myr (from birth to the present) is still in the main-sequence stage, which also means the impact of stellar wind loss on the cluster dynamical state can be ignored. 
The actual central mean positions (X, Y, Z) and orbital velocities (U, V, W) of the binary cluster in the Galactic tidal field are listed in Table~\ref{tab: Gal-para}. The metallicity of the binary cluster was set as $Z = 0.012$ \citep{2019JApA...40...33O} for stellar evolution.
In the simulation process, the binary cluster was set to evolve from the present day (cluster age at now $T_{\rm a}$ = 158~Myr) to 200~Myr (cluster age = 358~Myr), generating output snapshots with a time interval of 0.5~Myr, which included the Galactic positions and velocities of each star.

\begin{figure*}
   \centering
   \includegraphics[width=\textwidth, angle=0]{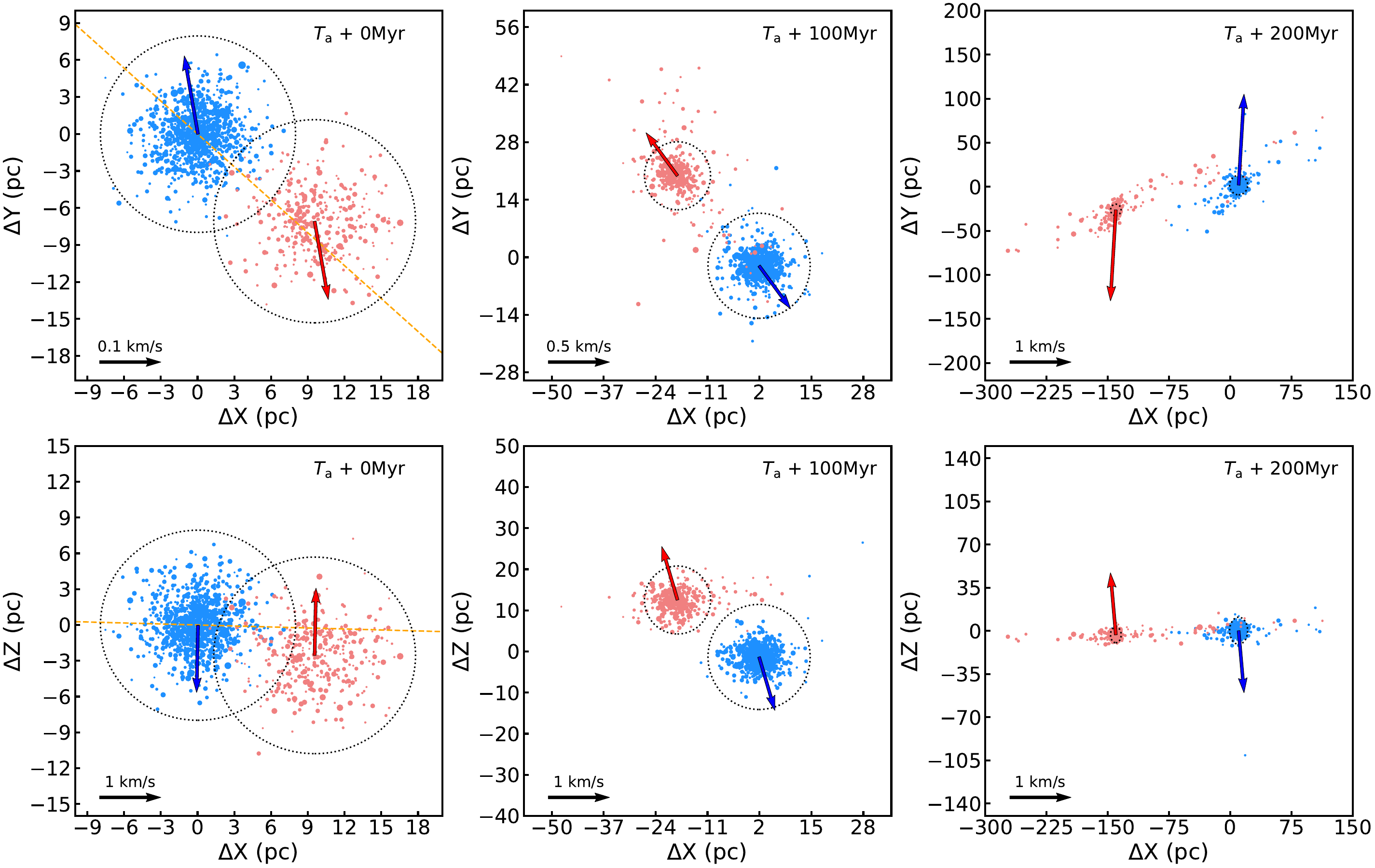}
   \caption{Future evolution of the two clusters (blue line in Figure~\ref{fig: orb}): snapshots at now (cluster age = $T_{\rm a}$ = 158~Myr), 100~Myr (cluster age = $T_{\rm a} + 100$~Myr), 200~Myr (cluster age = $T_{\rm a} + 200$~Myr) in the X–Y plane (top) and the X–Z plane (bottom) for the binary clusters with mass segregation degree $S$ = 0. The dodger blue and light coral points refer to the members of NGC 2323-a and NGC 2323-b, respectively, with blue and red arrows indicating relative velocities. The size of the points represents the stellar mass. The dotted and dashed circles present the tidal radius of NGC 2323-a and NGC 2323-b. We use the orange dashed lines to represent the line-of-sight direction.}
   \label{fig: nbody}  
\end{figure*}

Figure~\ref{fig: orb} presents the separation distance between the two clusters' centroids as a function of time. In summary, the future evolutionary process (blue and red lines) of the binary cluster can be divided into two stages. During the first 100~Myr, due to the distance between the two clusters being close enough, they rotate with each other, and their separation distance oscillates with a variation of less than 50~pc. The two clusters undoubtedly constitute a binary cluster and exhibit interaction during this dynamic stage. At 45~Myr, the minimum distance between cluster centroids is only 8.6~pc. It is noticed that, because of the gravitational interaction between the two clusters, the small mass cluster (NGC 2323-b) gradually expands after the close encounter occurs at 45~Myr, resulting in a change in the mass of NGC 2323-b. At the time of 100~Myr, since NGC 2323-b will expand into a looser cluster (approximately 20\% of stars are stripped outside the tidal radius of the cluster), the two clusters lack sufficient mass to remain gravitationally bound to each other, causing them to drift apart gradually. Then, from 100~Myr to 200~Myr, the separation distance between the two clusters increases larger and larger, promoting them to evolve independently along their orbits in the Milky Way, ultimately becoming two isolated star clusters separated by hundreds of parsecs. We also found that the influence of mass segregation ($S$ = 1) on the dynamic evolution of the binary cluster is feeble for cluster age $<$ 200~Myr: the presence or absence of mass segregation does not affect the separation distance of the cluster. 

Overall, the evolution of the binary cluster with their mean properties can be divided into two different dynamical states: one is the phase of being bound by the cluster's gravitational potential, and the two clusters evolved together as a binary cluster to undergo dynamical interaction; the other is the orbital motion that occurs after breaking free from the gravitational constraints between each other, being solely influenced by the gravitational potential of the Milky Way. The binary cluster evolves into two isolated open clusters in a few hundred million years. 

Moreover, we repeated the backward and forward simulations 500 times with Markov Chain Monte Carlo (MCMC) \citep{1986mcm..book.....K,10.1214/aos/1176325750} sampling over the measured uncertainties in the 3D positions and velocities as shown in Table~\ref{tab: para}. The numerical simulations with different sampling parameters are displayed in Figure~\ref{fig: orb}. Figure~\ref{fig: sep-hist} exhibits the orbital separations at birth time and 200~Myr, which states many possibilities of the origin and future evolution of the two clusters. We can see that the orbital separations for both backward and forward simulations are not very robust, demonstrating that the $N$-body simulation of the two clusters is sensitive to the initial parameters. The median values of the orbital separations are about 300~pc at birthtime and 260~pc at 200~Myr. We used the scipy package\footnote{\url{https://docs.scipy.org/doc/scipy/reference/generated/scipy.stats.gamma.html}} with gamma function to fit the three probability density distributions in Figure~\ref{fig: sep-hist}, respectively. The mode values we obtained from the gamma fitting are $75.1\pm26.2$~pc, $60.9\pm23.6$~pc, and $59.4\pm23.8$~pc.

To investigate the interaction and morphological characteristic of the two clusters with the mean property parameters, we plot the snapshots at different times in Figure~\ref{fig: nbody}, including 0~Myr ($T_{\rm a} + 0$~Myr), 100~Myr ($T_{\rm a} + 100$~Myr), and 200~Myr ($T_{\rm a} + 200$~Myr), where $T_{\rm a}$ is the cluster age at present time ($T_{\rm a}$ = 158~Myr). The snapshot of 0~Myr is consistent with the observational data, presenting the morphological characteristic of a typical star cluster. The separation distance between the cluster centroids is about 12.3~pc at present-day, and the relative velocity direction of the two clusters is parallel, indicating that the interaction between them is not strong, so it is unlikely that they will eventually merge into one cluster. At the snapshot of 100~Myr, the less massive cluster NGC 2323-b gradually expands after flying over another cluster NGC 2323-a several times. As the expansion of the cluster NGC 2323-b, its mass experiences a moderate loss, leading to the end of the interacting binary cluster stage. Eventually, when the two clusters evolve to the time of 200~Myr, they happen to complete one orbital period in the Milky Way. However, they will become two disconnected single clusters, with significant tidal tails extending to about 300~pc and a separation distance reaching 150~pc.

\subsection{Tidal tail}\label{sub: tt}

\begin{figure*}
   \centering
   \includegraphics[width=\textwidth, angle=0]{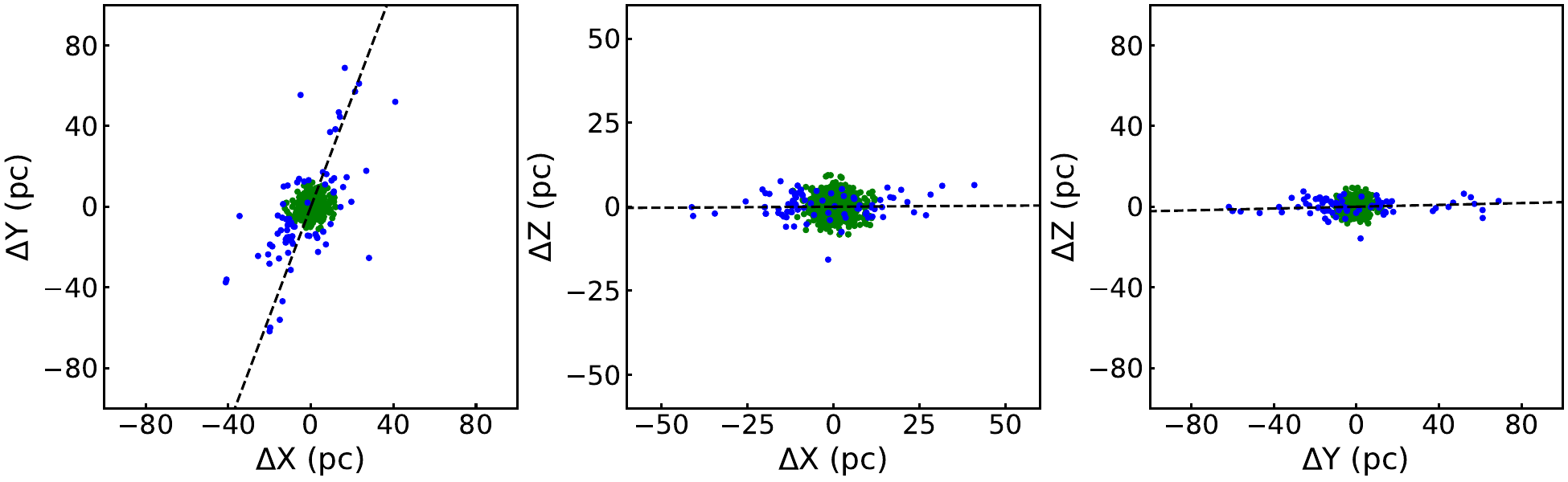}
   \caption{The ($\Delta$X-$\Delta$Y, $\Delta$X-$\Delta$Z, $\Delta$Y-$\Delta$Z) distributions of the members for the simulated `single cluster' NGC 2323 at present-day in Galactic coordinates. The green and blue points refer to the stars inside and outside the cluster tidal radius ($r_{\mathrm{tid}} = R (M_{\mathrm{oc}}/3M_{\mathrm{MW}})^{1/3}$ \citep{1962AJ.....67..471K,2022MNRAS.517.3613K}, where $M_{\mathrm{oc}}$ and $M_{\mathrm{MW}}$ refer to the masses of the cluster and the Milky Way, and $R$ represents the distance of the cluster from the Galactic center). We used black dashed lines to present the line-of-sight direction. $\Delta$X, $\Delta$Y, and $\Delta$Z are calculated relative to the cluster center.}
   \label{fig: single-OC}  
\end{figure*}

\begin{figure*}
   \centering
   \includegraphics[width=0.8\textwidth, angle=0]{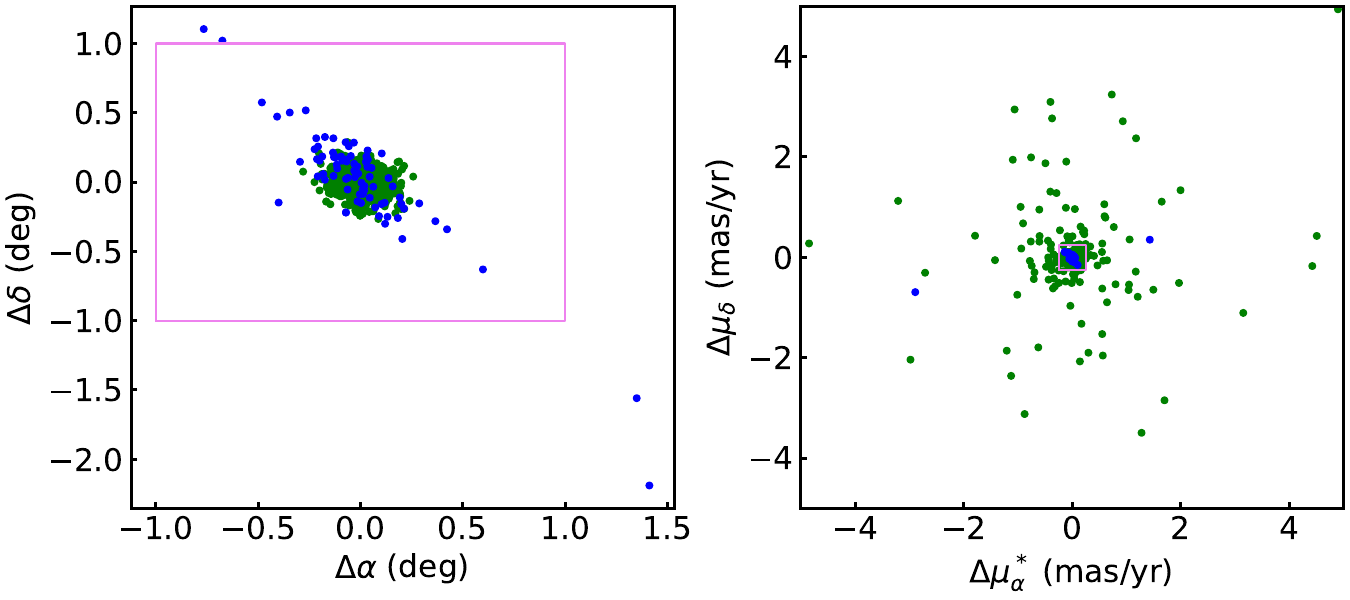}
   \includegraphics[width=0.8\textwidth, angle=0]{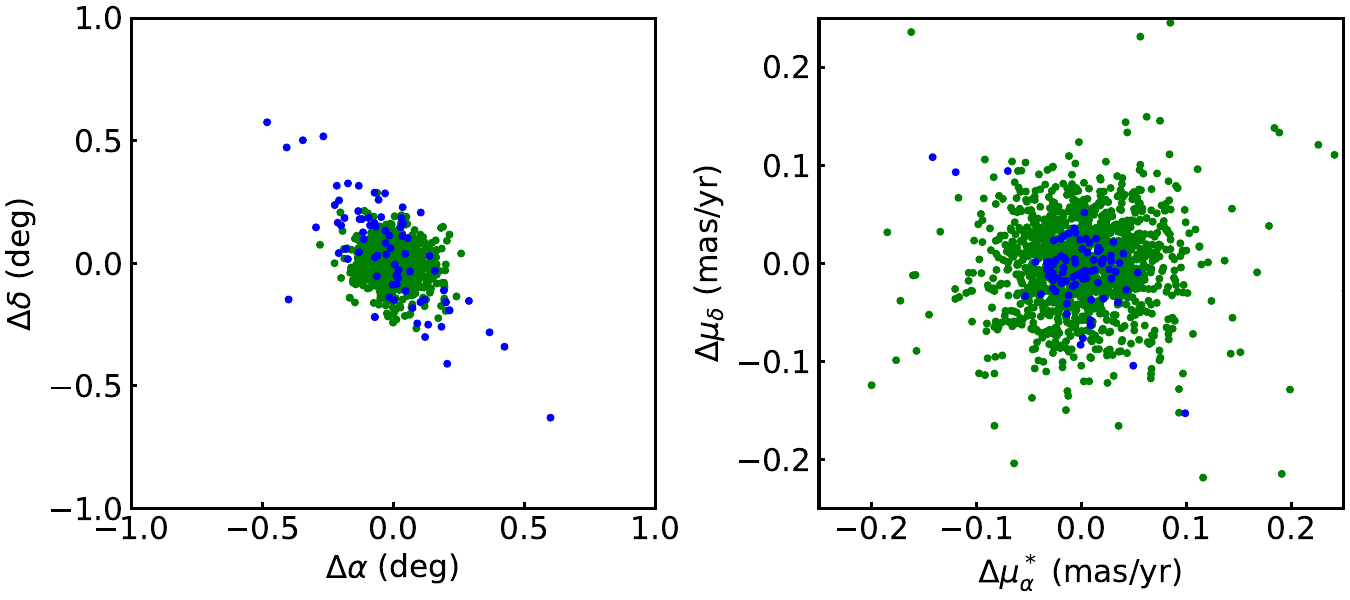}
   \caption{The spatial ($\Delta \alpha$, $\Delta \delta$) and proper motion ($\Delta \mu_{\alpha}^*$, $\Delta \mu_{\delta}$) distributions of the simulated `single cluster' NGC 2323 at present-day. The green and blue points correspond to those in Figure~\ref{fig: single-OC}. $\Delta \alpha$, $\Delta \delta$, $\Delta \mu_{\alpha}^*$, and $\Delta \mu_{\delta}$ are calculated relative to the cluster center. The bottom panels are the magnified views within the boxes in the top two panels.}
   \label{fig: single-OC-pm}  
\end{figure*}

It is well known that the tidal tails of open clusters may be created by the tidal forces from the Milky Way potential or a nearby massive object \citep{2019A&A...621L...2R,2020ApJ...889...99Z,2021A&A...647A.137J,2021A&A...655A..71W,2022MNRAS.514.3579B,2022RAA....22e5022B}. It is worth mentioning that \citet{2022RAA....22e5022B} documented a tidal structure stretching to around 270~pc of the nearby cluster COIN-Gaia 13 (cluster age 250 Myr), while \citet{2022MNRAS.514.3579B} similarly observed a 260~pc long tidal tail of the open cluster NGC 752 (cluster age 1-2~Gyr). These tidal tails are mainly torn by the differential rotation induced by the Galactic potential. Similarly, the less massive `cluster' in this work is possibly to be the tidal tail of NGC~2323. Here, we employed $N$-body simulation to check if this single cluster NGC~2323 could have developed tidal tails in the Milky Way since its birth.

We assumed that NGC~2323 was born as an isolated cluster and used the similar approach described in Section~\ref{sub: bc} to mock this `single cluster' at birth time with the McLuster package \citep{2011MNRAS.417.2300K}. The initial state of the cluster is set as the simulated initial state of the cluster 158~Myr ago from backward simulation. To check if NGC 2323 would develop tidal tails after 158~Myr, we simulated its dynamical evolution in the Milky Way from the birth time to present-day with PETAR \citep{2020MNRAS.497..536W}. The position and proper motion distributions at present-day are presented in Figure~\ref{fig: single-OC} and Figure~\ref{fig: single-OC-pm}. This cluster forms a tidal tail (the blue points), however, we can see that the proper motion distribution at present-day exhibits a single concentration, which is different from what we observed with {\it Gaia} DR3. It is noted that some stars, color-coded as green, have larger relative velocities, which are caused by the orbital motion of binary stars. Because the 1 sigma upper age of this cluster from isochrone fitting is 281~Myr in Section~\ref{sec: ana}, we set the simulation time as 281~Myr. However, there are still no two concentrations in the proper motion distribution of the tail and core members at 281~Myr from the $N$-body simulation. The two distinct clumps or concentrations in the proper motion distribution in Figure~\ref{fig: member} demonstrate that the less massive `cluster' is unlikely to be the tidal tails of this young open cluster NGC 2323, which are torn by the Galactic potential. It is worth mentioning that the incompleteness of {\it Gaia} data especially for the faint stars impacts what we can see from the observational data. For this cluster, we assumed that the members are complete for sources brighter than 18~mag by checking the {\it Gaia} selection function \citep{2023A&A...669A..55C} in this region and included the unseen stars in the simulation by integrating the extrapolated observed mass function down to the lower mass limit (0.08~$\rm M_\odot$). This incompleteness, along with factors such as extinction and binary stars, would lead to the inconsistency between observational and simulation data. As well, we can not exclude the possibility that the tidal tails may be caused by a nearby massive object, and further numerical simulations are needed to investigate this in the future.

\section{Summary} \label{sec: sum}

Based on the spatial and proper motion distributions provided by {\it Gaia} DR3, we discovered that the well-known open cluster NGC 2323 has a binary structure, which might be classified as a binary cluster containing two individual components, NGC 2323-a and NGC 2323-b, or a cluster with tidal tails created by a nearby massive object. Further investigation of the cluster properties shows that the two `clusters' are located very close to each other (three-dimensional $\Delta \rm pos = 12.3$~pc) and have similar yet discernible kinematics (two-dimensional $\Delta \rm V = 2.2$~km s$^{-1}$) and the same age (158~Myr), which indicates they may have a common origin and currently in physical interaction. 

The close position and similar velocity of the two structures suggest that NGC~2323 is probably a binary cluster. To obtain a comprehensive understanding of the formation and evolution of this binary cluster, we employ the PETAR to trace back their birthplace and deduce their dynamical evolutionary fate. Our work shows that the two clusters may be born in primordial pairs with a separation of about 27.2~pc and then orbit each other as a binary cluster. In the future, undergoing multiple flyovers with each other, the smaller cluster (NGC2323-b) will gradually expand and lose some of its stellar mass. Then, about 300~Myr later, the two constituent parts of the binary cluster may leave apart and eventually evolve into two independent clusters.

The $N$-body simulation indicates that the less massive cluster is unlikely to be the cluster tidal tails torn by the differential rotation of the Milky Way. However, this tail may be created by a nearby massive object. More future simulations are needed to further investigate this subject.

\section*{Acknowledgments}
We express our gratitude to the anonymous referee for their valuable comments and suggestions, which are very helpful in improving our manuscript.
This work is supported by the National Natural Science Foundation of China (NSFC) through grants 12090040, 12090042, 12073060, 21BAA00619, 12073090, and 12233013. 
J.Z. would like to acknowledge the Youth Innovation Promotion Association CAS, the science research grants from the China Manned Space Project with NO. CMS-CSST-2021-A08, the Shanghai Science and Technology Program (20590760800), and the Science and Technology Commission of Shanghai Municipality (Grant No. 22dz1202400). L.W. thanks the one-hundred-talent project of Sun Yat-sen University, the Fundamental Research Funds for the Central Universities, Sun Yat-sen University (22hytd09). K.W. would like to acknowledge the Postgraduate Research Scholarship (grant PGRS1906010) of Xi'an Jiaotong-Liverpool University (XJTLU). 
S. Q. acknowledges the financial support provided by the China Scholarship Council program (Grant No. 202304910547). 
This work was partially funded by the Spanish MICIN/AEI/10.13039/501100011033 and by the ``ERDF A way of making Europe" funds by the European Union through grant PID2021-122842OB-C21, and the Institute of Cosmos Sciences University of Barcelona (ICCUB, Unidad de Excelencia `Mar\'{\i}a de Maeztu’) through grant CEX2019-000918-M. FA acknowledges financial support from MCIN/AEI/10.13039/501100011033 and European Union NextGenerationEU/PRTR through grant RYC2021-031638-I.

This work has made use of data from the European Space Agency (ESA) mission {\it Gaia} (\url{https://www.cosmos.esa.int/gaia}), processed by the {\it Gaia} Data Processing and Analysis Consortium (DPAC, \url{https://www.cosmos.esa.int/web/gaia/dpac/consortium}). Funding for the DPAC has been provided by national institutions, in particular, the institutions participating in the {\it Gaia} Multilateral Agreement.


\bibliography{sample631}{}
\bibliographystyle{aasjournal}

\end{CJK*}
\end{document}